\documentclass[acmsmall]{acmart}

\AtBeginDocument{%
  \providecommand\BibTeX{{%
    \normalfont B\kern-0.5em{\scshape i\kern-0.25em b}\kern-0.8em\TeX}}}

\usepackage{natbib}
\usepackage{url}

\acmBooktitle{}
\begin{document}
\title{DiPD: Disruptive event Prediction Dataset from Twitter}

\author{Sanskar Soni}
\email{2018ucp1265@mnit.ac.in }

\author{Dev Mehta}
\email{2018ucp1382@mnit.ac.in }

\author{Vinush Vishwanath}
\email{2018ucp1491@mnit.ac.in }

\author{Aditi Seetha}
\email{aditiseetha28@gmail.com}

\author{Satyendra Singh Chouhan}
\email{sschouhan@mnit.ac.in}

\affiliation{%
  \institution{Malaviya National Institute of Technology Jaipur}
  \streetaddress{JLN Marg}
  \city{Jaipur}
  \state{Rajasthan}
  \country{INDIA}
  \postcode{302017}
}








\begin{abstract}

Riots and protests, if gone out of control, can cause havoc in a country. We have seen examples of this, such as the BLM movement, climate strikes, CAA Movement, and many more, which caused disruption to a large extent. Our motive behind creating this dataset was to use it to develop machine learning systems that can give its users insight into the trending events going on and alert them about the events that could lead to disruption in the nation. If any event starts going out of control, it can be handled and mitigated by monitoring it before the matter escalates. 
This dataset collects tweets of past or ongoing events known to have caused disruption and labels these tweets as 1. We also collect tweets that are considered non-eventful and label them as 0 so that they can also be used to train a classification system.
The dataset contains $94855$ records of unique events and $168706$ records of unique non-events, thus giving the total dataset $263561$ records. We extract multiple features from the tweets, such as the user's follower count and the user's location, to understand the impact and reach of the tweets. 
 This dataset might be useful in various event related machine learning problems such as event classification, event recognition, and so on.

\end{abstract}






\maketitle
\section{Introduction and Specification}
A disruptive event is an event that obstructs routine process to fulfil their goal and instructed by many unauthorized
sources~\cite{alsaedi2015feature,alsaedi2017can}. Its duration is often unpredictable and  it may happen for one or two days or may be continuing for several days. It can spoil law and order situations that may lead to a civil unrest~\cite{panagiotopoulos20125,bahrami2018twitter}. The objective of such events are often unclear
therefore it happens in a very unplanned and unstructured manner. 
Nowadays social media has become a primary source of information. The common man and authorities report every event or incident on social media. For example, social media became a tool for Protest against Citizen
Amendment Act (CAA) in India~\cite{web2}. In such scenarios, any misleading information spread over the social
microblogging sites can convert peaceful protest into violence~\cite{web1}. However, if we can get the early indication of the
disruptive events using social media information then preventive measures can be taken at an early stage. In this work, first, we are collecting Disruptive Event data from
the social media (twitter) that will be updated and gathered continuously. A part of this dataset is now being available and published. Table~\ref{specification} shows the specification of dataset. The description of the dataset is given in the subsequent sections.

\begin{table}[htb]
\caption{Specifications Table}
\centering

\label{specification}
\begin{tabular}{p{4.5cm}p{8.0cm}}
\hline 
Subject Area                   & Computer science                                                                                                                                                                                                                                                                                                                                                                                   \\
Specific subject Area          & Artificial Intelligence                                                                                                                                                                                                                                                                                                                                                                            \\
Type of data                   & Textual tweets along with numeric attributes                                                                                                                                                                                                                                                                                                                                                       \\
How data was acquired          & Data were acquired by extracting tweets using the open source tweepy library along with the features of that tweet                                                                                                                                                                                                                                                                                 \\
Data Format                    & Raw csv file                                                                                                                                                                                                                                                                                                                                                                                       \\
Parameters for data collection & Only those tweets having keywords from the events and the non event set were extracted.                                                                                                                                                                                                                                                                                                            \\
Description of data collection & The data ( a total of 2 files belonging to the event and non-event categories)comprises of features extracted from the tweeter's profile itself using the Tweepy API. The data consists of 6 features, namely the time at which it was tweeted, retweet count, follower count, location(only of those whose location services were turned on at the time of posting), username, and statuses count \\
Data source location           & Worldwide                                                                                                                                                                                                                                                                                                                                                                                          \\
Data accessibility             & The dataset can be accessed through the URL: \textit{\url{https://vj-creation.github.io/krl-webpage/resources.html}}  or GitHub Link
\textit{\url{https://github.com/devmehta01/DiPD}}
                                                                                     
\\
Related research article       &          Given in the references     

\\

\hline
\end{tabular}
\end{table}

\section{Value of the Data}

\begin{itemize}
    \item This data consists of a collection of eventful and non-eventful tweets. Each tweet is assigned a value of 1 or 0. 1 meaning eventful and 0 meaning non-eventful. The data can be used as input for machine learning systems.
    
    \item Machine Learning researchers can benefit from this dataset, while Governmental and Security agencies can also benefit from the machine learning models resulting from this dataset. Government organizations can use this on future tweets to keep track of events and mitigate them before they become violent.
    
    \item Features such as tweet location are extracted and can be used to determine where the events are occurring. It also includes features such as user followers and retweet count, which can be used to find the impact factor of the tweet.

    \item The provided dataset can also be used as a performance benchmark for developing state-of-the-art machine learning systems for disruptive event prediction

\end{itemize}

\section{Data Description}
This paper contains twitter data for the prior prediction of disruptive events. The target class contains two labels - event and non-event. The dataset contains 7 attributes and 263,561 records, out of which 94,855 records are of event class, and 168,706 records are of non-event class. The attributes described in Table~\ref{attributes}, contain details about the tweet and information about the user. The data contains numerical and continuous data to be used for analysis based on classification, prediction, segmentation, and association algorithms. The dataset folder contains four csv files, two for event records(containing both raw and preprocessed tweets) and another for non-event records (with raw and preprocessed tweets).

\begin{table}[htb]
\caption{Attributes Table}
\centering

\label{attributes}
\begin{tabular}{lp{2.0cm}p{6.8cm}p{2cm}p{1cm}}
\hline 
Nr. & Attribute       & Description                                                                            & Format                & Values    \\ \hline
1.  & created\_at     & This feature shows us the time and date at which tweet was tweeted                     & YYYY/MM/DD   HH:MM:SS &           \\
2.  & retweet\_count  & This feature depicts the number of time the given tweet was retweeted                  & Numeric               &           \\
3.  & follower\_count & This feature shows the number of followers the tweeter has                             & Numeric               &           \\
4.  & location        & This feature depicts the approximate location of the place from where tweet was posted & Alphanumeric          &           \\
5.  & username        & The feature shows the twitter handle of the user                                       & Alphanumeric          &           \\
6.  & statuses\_count & The feature shows the number of tweets (including retweets) issued by the user         & Numeric               &           \\
7.  & label           & This feature shows whether the tweet lies in event or non-event category               & Boolean               & {[}0,1{]} \\ \hline
\end{tabular}
\end{table}

\subsection{Data Extraction}

In order to extract the tweets, Python's Twitter API 'Tweepy' was used. Event class data was gathered by using keywords containing examples of major disruptive events such as the Farmer's protests in India and the 'Black Lives Matter protests. Similarly, Non-event class data was obtained by using a different set of keywords. The algorithm avoids storing retweets. The attributes of the tweet stored are of two types: User-specific information such as the screen name and tweet-specific information such as the text, the number of retweets, the date and more. Extraction was performed multiple times over a few weeks so as to collect as many unique tweets as possible. The data was then preprocessed to remove duplicate tweets. The dataset is deliberately kept slightly unbalanced in accordance to the fact that most of the tweets are usually non-events.Few examples from the dataset are presented in Table~\ref{examples}.

\begin{table}[htb]
\caption{Example instances from the dataset}
\centering
\scriptsize
\label{examples}
\begin{tabular}{p{10.8cm}p{2.5cm}}
\hline 
Tweet & Label \\
\hline
@DschlopesIsBack @lickofcow @DoomerCoomer At a black lives matter protest. This isn't about self defense, this is about rittenhouse not liking BLM standing up to police brutality. Not liking people taking a knee, not liking the civil disturbances and protest. Thats why he killed them & 1(Event)     \\
\hline
@ArvindKejriwal Ideally pollution should stop after \#Diwali What is your excuse now?                                                                                                                                                                                                        & 1(Event)     \\
\hline
"*BABRI MASJID* The judgement given over the Babri Masjid verdict was utterly biased just make the majority happy!! \#BabriMasjidVictimOfInjustice https://t.co/qoX2z92ghS" 
& 1(Event)     \\
\hline
This affair has become a toxic combination of the methods of the \#MeToo movement and the escalating aggression by the United States ruling elite toward China.                                                                                                       & 1(Event)     \\
\hline
\#Bitcoin is at 63636 USD                                                                                                                                                                                                                                                                    & 0(Non-event) \\
\hline
Play the \#TheLastofUsPartll on grounded difficulty with permadeath enabled... With a PS4 that random ejects the disk! It's a different kinda rush cuz.                                                                                                                                      & 0(Non-event) \\
\hline
so glad seeing taeyong spending his time with his friends                                                                                                                                                                                                                                    & 0(Non-event) \\
\hline
Sorry Zomato we are not same.... Prefer Dal Bhaath or Kaanji Bhaath...... Your choices in vast India is very limited. @zomato @zomatocare https://t.co/btHnI91Rts                                                                                                                            & 0(Non-event)
\\
\hline

\hline
\end{tabular}
\end{table}

\begin{figure}[htb]
    \centering
    
  \includegraphics[height = 4.5 in, width = 5.0 in]{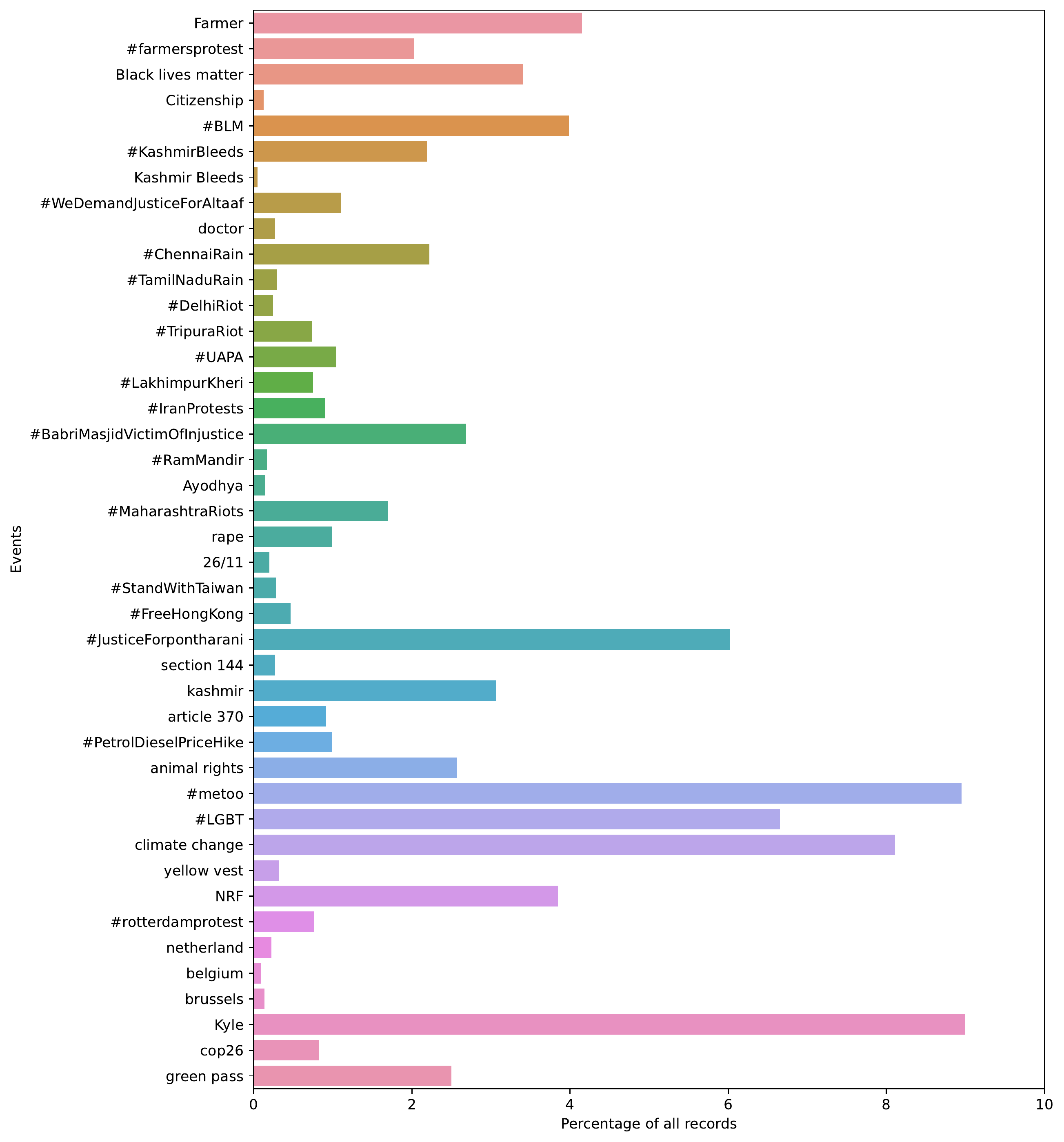}
    \caption{Percentage of tweets of various topics present in event category}
    \label{bar}
\end{figure}

\begin{figure}[htb]
    \centering
    
  \includegraphics[height = 1.8 in, width = 2.2 in]{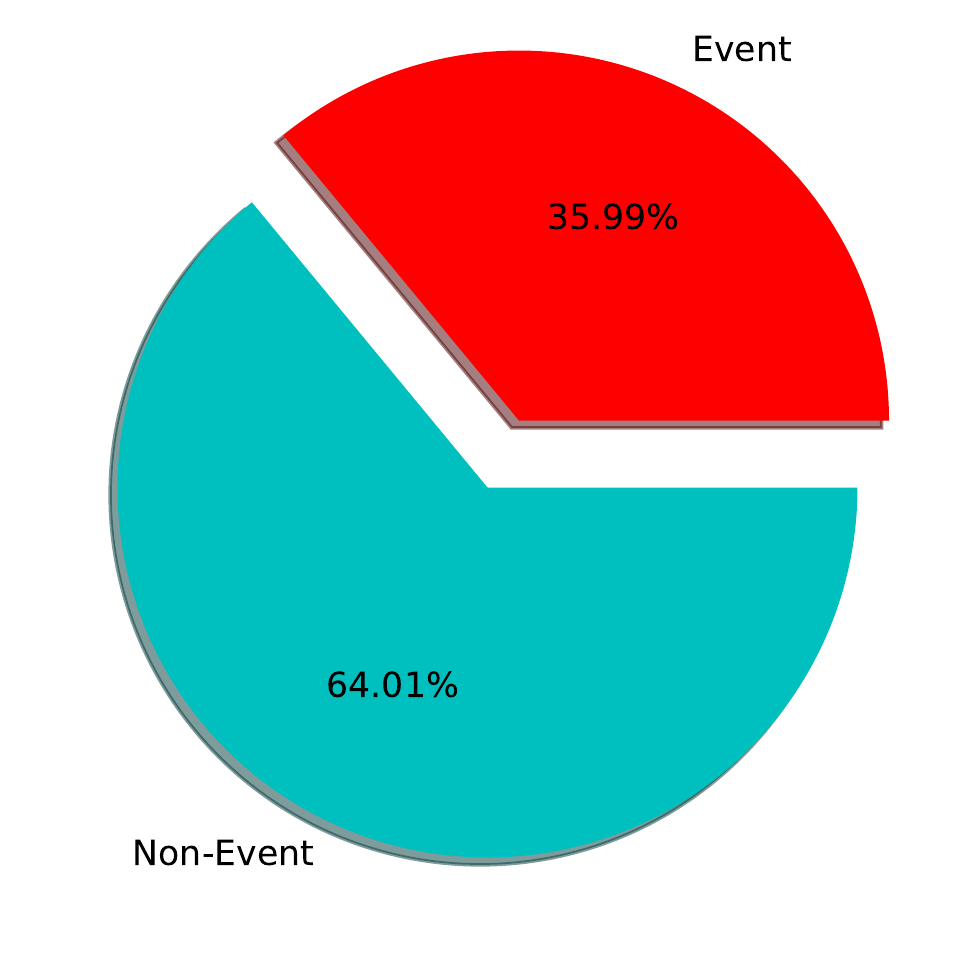}
    \caption{Proportion of eventful and non-eventful tweets present in dataset}
    \label{pie}
\end{figure}
\subsection{Distribution of tweets}
As illustrated in Figure~\ref{bar}, tweets from various countries and domains are extracted and their share in the whole dataset is presented. The share of different topics have been drwan out and one can clearly see the importance of \#metoo and climate change. The list of of event and non-event keywords which were used for extraction are given in Table~\ref{eventnonevent}.

\begin{table}[htb]
\caption{Example instances from the dataset}
\centering
\scriptsize
\label{eventnonevent}
\begin{tabular}{p{6.5cm}p{6.5cm}}
\hline 
Event Keywords & Non-Event Keywords \\
\hline
\#farmersprotest, Black lives matter, Citizenship, \#CAAProtest, \#BLM, \#KashmirBleeds, \#WeDemandJusticeForAltaaf, Nabanna  \#ChennaiRain, \#TamilNaduRain,  \#DelhiRiot , \#BangaloreRiot, \#TripuraRiot, \#UAPA, \#LakhimpurKheri, \#IranProtests, \#BabriMasjidVictimOfInjustice, \#RamMandir, Ayodhya, \#MaharashtraRiots, \#JusticeForNirbahaya, \#MaharashtraUnsafe4Women, rape, 26/11,  \#StandWithTaiwan, \#FreeHongKong, \#JusticeForpontharani,  section 144, kashmir , article 370, \#PetrolDieselPriceHike, animal rights, \#metoo, \#LGBT, climate change, yellow vest , NRF, \#rotterdamprotest, netherland ,belgium ,brussels ,Kyle Rittenhouse, cop26, green pass, & Bitcoin, Cricket, Football, Tennis, Clothes, Vacation, Crypto, Sports, Guitar, Keyboard, Happy Birthday, Movies, music, stocks, leisure, galaxy, NASA, Science, School,KPop,Fruits, Mango,Gym, Workout, Decoration, Mothers day, Teachers day, phone, violin, youtube, hiking, Exam, Haircut, Outfit,Diwali, Shopping, Spotify, Facebook, Samsung, Apple, Phone, Marvel, DC, Pizza\\

\hline
\end{tabular}
\end{table}

The distribution as a whole of event and non-event classes has been shown in the Figure~\ref{pie}. About 36\% belong to the event class and the rest to the non-event category. Even though the keywords for event category were more but the tweets extracted were less.

\section*{Acknowledgment}

This work is supported by a Research Grant under National Supercomputing Mission (India), Grant number: \textit{DST/NSM/R\&D\_HPC\_Applications/2021/24}.

\bibliographystyle{ACM-Reference-Format}
\bibliography{main}

\end{document}